 \documentclass[nohyper,12pt,letterpaper]{JHEP}
 \usepackage{epsfig}
 
 \def \eqn#1#2{\begin{equation}#2\label{#1}\end{equation}}

 \title{Entropy and initial conditions in cosmology}
 \author{T.\,Banks\\
 Department of Physics and SCIPP\\
 University of California, Santa Cruz, CA 95064\\
 E-mail: \email{banks@scipp.ucsc.edu}\\
 {\it and}\\
 Department of Physics and NHETC, Rutgers University\\
 Piscataway, NJ 08540}

 \abstract{I discuss the Boltzmann-Penrose question of why the initial conditions
 for cosmology have low entropy.  The modern version of Boltzmann's answer to this
 question, due to Dyson, Kleban and Susskind, seems to imply that
 the typical intelligent observer arises through thermal
 fluctuation, rather than cosmology and evolution.   I investigate
 whether this can be resolved within the string landscape.   I end
 with a review of the suggestion that Holographic Cosmology provides a simpler
 answer to the problem. This paper is a revision of unpublished work from the spring of
 2006, combined with my talk at the Madrid conference on String theory and Cosmology, Nov 2006.}
 \received{} \accepted{}
 \preprint{hep-th{/0701146}\\\\ \\}
 
\begin{document}

 \section{\bf Introduction}

 The Second Law of Thermodynamics is one of our most robust and
profound physical principles.   Its origins in the combination of
the laws of mechanics and probability were, in large measure,
understood in the 19th century, though quantum mechanics simplifies
some of the considerations.   While a completely rigorous derivation
still eludes us, the modern physicist's understanding of the way in
which this principle should be applied to laboratory and
astrophysical systems is adequate for almost all purposes.

Ever since the discovery of the Second Law, physicists have been
faced with the question of why the universe began in a low entropy
state, so that we can see the Second Law in operation. Roger Penrose
has emphasized this question in the context of modern
cosmology\cite{penrose} .   He has also rejected the claims of
inflation theorists that the problem is resolved or avoided by
inflation.   I agree with Penrose on this point and will discuss it
further below.

The increase in entropy that we observe in laboratory systems is a
consequence of fluctuations out of equilibrium (sometimes forced on
the system by actions of the experimenter).   A system in
equilibrium has already maximized its entropy and sees no further
entropy increase until a fluctuation puts it into a low entropy
state.  Penrose has emphasized for many years, that the questions of
why the universe began in a low entropy state, and what the nature
of that state is are among the most important puzzles in cosmology.
It is believed by many physicists, though I have not been able to
find a precise reference, that Boltzmann himself proposed that the
answer to this question is the same as that for systems in the
laboratory\cite{vilenkinquote}. The universe is a finite system and
the low entropy beginning was just one of the inevitable random
fluctuations that any finite system undergoes.

A modern version of Boltzmann's explanation was recently proposed as
a straw man in a paper by Dyson, Kleban and Susskind\cite{DKS}
(DKS). The main purpose of these authors was to criticize the
proposal that the origin of low entropy initial conditions is a
fluctuation. I will review their work extensively below and
criticize their conclusions about what a typical history in their
model looks like. My criticism depends on a particular point of view
about quantum gravity, which I will detail in the appropriate place.
However, I will show that although the DKS model does not suffer
from the problems described by its authors, it does suffer from the
problem of Boltzmann's Brain (BB)\footnote{This was pointed out to
me by R. Bousso.}. That is, if we make the hypothesis that
intelligent observers are isolated physical systems, whose state can
be described by a very small subset of the degrees of freedom of the
universe, then, in the DKS model,  it is much more probable to
produce a single intelligent observer by thermal fluctuation, than
it is to observe the cosmological evolution they propose as the
history of the universe. We will call such thermally produced
intelligences, {\it Boltzmann Brains}.  Here are references to the
recent rash of papers on this subject\cite{bbrains}.

The String Landscape proposes an explanation of the part of the
history of the universe that we can observe, which is very similar
to that of the DKS model.   It will also contain the phenomenon of
thermally produced intelligence, unless all meta-stable dS states in
the Landscape, which are capable of supporting life,\footnote{There
is not the slightest possibility that we will be able to determine
whether a given model of particle physics supports the kind of
complex organization that we call intelligence, unless the low
energy effective Lagrangian at nuclear physics scales is very close
to that of the standard model.   In this paper life will always mean
the kind of life we know exists in the real world.}, decay rapidly
into negative cosmological constant Big Crunches. The Landscape
differs from the DKS model in that it does not really give an
explanation of how the universe gets into the low entropy state from
which it tunnels into the basin of attraction in which we find
ourselves. Thus, at its present state of development, it does not
address the question posed by the Second Law. Nonetheless, many
landscape enthusiasts are nervous about the possibility that
Boltzmann's Brain, isolated in a spaceship, is the typical form that
intelligence takes in the Landscape.  Indeed the results of
\cite{dinelindeshenkeretal} suggest that the landscape contains many
approximately supersymmetric dS states, whose decay is extremely
slow.   Certain methods of counting probabilities in the landscape
tell us that the landscape is dominated by the progeny of such
states.   In order to completely remove the spectre of BB's without
a deeper understanding of initial conditions in the landscape, one
would have to show that {\it none} of these nearly SUSic states has
low energy physics like that of our own world.

After discussing these abstruse questions, and counting how many
angels there are on a pin head (a p-brane?) , I will take up another
explanation of low entropy initial conditions.  Holographic
cosmology was invented\cite{holocosm} in order to find a proper
description of the Big Bang.   It was based on the observation of
Fischler and Susskind\cite{fishsuss} that the only FRW cosmology
compatible with the holographic principle at very early times has
equation of state $p =\rho$\footnote{Generic cosmologies with less
stiff matter, violate the holographic principle prior to some time
$t_0$.   It is interesting that given the current conditions in our
universe, backward extrapolation of ordinary cosmology gives $t_0$
of order the Planck time. }. Assuming an extensive entropy, this
implies that the entropy density $\sigma \propto \sqrt{\rho}$, which
is what one obtains from a collection of black holes at distances of
order their Schwarzschild radii.   In order for this configuration
to remain steady in an expanding universe, the black holes have to
continually merge so that each horizon volume is filled with a black
hole.   We called this the {\it dense black hole fluid}. It is not
clear what one means by such a configuration in general relativity.
However, in \cite{holocosmmath} the authors constructed an explicit
quantum mechanical model, satisfying a set of plausible rules for
quantum cosmology, and obeying the scaling laws of the flat FRW
cosmology with $p=\rho$. The space-time geometry was an emergent
quantity, derived from the rules of the quantum system.  By
contrast, space-time topology was fixed by an {\it a priori}
lattice. The model also had the properties one expects of a dense
black hole fluid.   The entropy in each horizon volume was maximized
and the energy was that of a black hole filling the horizon.

In the penultimate section of this paper, I will describe
holographic cosmology in a bit more detail, and indicate how it
addresses the problem of low entropy initial conditions. In the
conclusions I will compare the various approaches to this problem
and suggest avenues for further progress.

\section{\bf DKS {\it redux}}
\subsection{The DKS model}

As noted above, when we apply the Second Law of Thermodynamics to
cosmology, we encounter a puzzle. Much folklore among inflation
theorists is devoted to the claim that inflation resolves this
problem and simultaneously hides the precise nature of the initial
state from our view.  Penrose rejects this claim, and I agree with
him, though my reasons do not coincide with his.   The initial
inflationary patch contains only a small number of degrees of
freedom that can be described by effective field theory\footnote{To
be more precise, we should talk about a causal diamond of an
observer in the initial inflationary patch, and apply the covariant
entropy bound\cite{fsb} .}.   Most of the degrees of freedom of the
observable universe are not well described by quantum field theory
until a large number of e-folds occur.   The most sophisticated
argument for the conventional discussion of these degrees of freedom
starts from the {\it assumption} that they were in the ground state
of some slowly varying Hamiltonian, which approaches the
conventional Hamiltonian of field theory in the inflationary
background, co-moving mode by mode, as the physical size of each
mode crosses the Planck scale. There are many ad hoc assumptions in
this treatment, but it is clear that the low entropy initial
condition is {\it put in} by assuming that the system was in its
ground state.  The excited states of every large quantum system that
we know, are highly degenerate, and the adiabatic theorem simply
does not apply to generic initial conditions chosen as a linear
combination of highly degenerate states.  Thus, most conventional
discussions of inflation assume a very special state for a huge
number of degrees of freedom at a time when we do not have a
reliable dynamical description of these variables.   In this sense,
inflation {\it does not} solve the problem of homogeneity and
isotropy of the early universe.

A few years ago, DKS \cite{DKS} proposed another approach to the
problem of low entropy initial conditions. It was based almost
entirely on semi-classical notions, sprinkled with a few grains of
holographic wisdom.   These authors concluded that their model
predicts phenomena, which are not verified observationally (we will
review their arguments below).  Indeed the main intent of their
paper was to show that the hypothesis that the beginning of the
universe is a fluctuation in a finite entropy system was
incompatible with observation. The purpose of the present section is
to show that a proper understanding of the quantum dynamics of de
Sitter space, removes the objections of Dyson and her collaborators.
Indeed, I would claim that, apart from the problem of Boltzmann
Brains, their model presents us with a solution of the problem of
initial conditions, which can be investigated with semi-classical
tools.

The basic idea of the proposal of \cite{DKS} was that the universe
we live in is a de Sitter space with c.c. $\Lambda$.   This system
has finite entropy, and thus Poincare recurrences.   DKS proposed
that the cosmology we observe is the result of such a recurrence, in
which the universe entered a very low entropy state.   For reasons
which will become apparent, I will discuss a particular version of
this model, which has a meta-stable, low entropy excitation modeled
by a subsidiary minimum of the effective potential with c.c.
$\Lambda_1 \gg \Lambda$\footnote{Although this version was not
emphasized in the DKS paper, it featured prominently in
conversations I had with Lenny Susskind, in which he explained the
DKS model.}. The process of tunneling to the low entropy minimum and
back is well described in the semiclassical approximation, by a
Coleman De Luccia (CDL)\cite{cdl} instanton. The tunneling rates
calculated from this instanton indeed satisfy the law of detailed
balance\cite{heretics} and are a compelling piece of evidence for a
model of quantum de Sitter space as a finite system with Poincare
recurrences\footnote{It also provides evidence for the spectrum of
the static dS Hamiltonian that I will describe below.   The law of
detailed balance involves entropies rather than free energies, which
makes sense if all of the states are at energies below the dS
temperature.}.

The quantum mechanics is supposed to be that of a system of a finite
number of states.  It has a static Hamiltonian, $H$, which is the
quantum representation of the static time-like Killing vector of a
given observer's causal diamond. All of the eigenstates of $H$
resemble the lower dS vacuum macroscopically. Particle excitations
of the lower minimum are assumed to be identical to the low energy
particle spectrum\footnote{Here spectrum means the spectrum of the
operator $P_0$, to be introduced below, not that of the operator
$H$.} in the real world (so there are other low energy fields
coupled to the inflaton). One way to get a universe that resembles
our own is to have a fluctuation into the very low entropy false
vacuum state, followed by rapid (on the time scale of the true
vacuum Hubble constant), Coleman-De Luccia (CDL) tunneling back to
the true vacuum. \cite{klebanetal} argue that after such a tunneling
event, the only way to get galaxies is to have a sufficiently flat
potential in the basin of attraction of the true minimum, so that
one gets at least $58$ e-folds of inflation. Field theoretic
naturalness suggests that we are unlikely to have many more e-folds
than this\cite{tbetal}, and that the inflaton potential is of the
form $\mu^4 v(\phi / m_P)$, with Planck scale couplings to ordinary
matter. For potentials of this form, the normalization of primordial
inflationary fluctuations scales like $Q \sim \mu^2 / m_P^2$.   The
reheat temperature is of order $\mu^3 /m_P^2$ and the growth of the
scale factor between inflation and reheating is $(m_P /
\mu)^{8\over3}$.  We must have $\mu \ll m_P$ in order to use this
effective field theory description, while if $\mu$ is too small the
universe will collapse into inhomogeneous clumps of Bose condensate
(or black holes) because many scales will go non-linear before
radiation domination begins.   The value of $\mu$ that reproduces
the observed CMB fluctuations avoids both of these disasters.

This version of the DKS model thus invokes both old and slow roll
inflation. In doing so, it avoids the criticism of generic inflation
models that we reviewed above.   The period spent in the meta-stable
state with c.c. $\Lambda_1 \gg \Lambda$ explains why we begin with
low entropy and also why the initial conditions for cosmology are
homogeneous and isotropic.

It is thus plausible that a model like this\footnote{With the value
of $\mu$ either determined by a unique microphysical theory or
chosen from a landscape by the sort of environmental selection
adumbrated above.} can reproduce a universe like our own. The
authors of \cite{DKS}do not dispute this.  Rather, they suggest that
as we cycle through the infinite set of Poincare
recurrences\footnote{In a previous paper\cite{nightmare} we
criticized this prediction because it involves properties of the
mathematical universe, which can never be tested {\it in principle}.
According to DKS, their mathematical model contains one history that
resembles our universe and predicts everything we see, but also
makes predictions about events that are in principle unobservable,
to any observer which experiences that history. If one takes the
point of view that the role of theoretical physics is simply to make
mathematical models that explain and predict what we can observe,
then there is little difference between this situation and one in
which we have many different mathematical models, one of which
describes what we see, while the others don't. In the current paper,
I accept the {\it assumption} that one must average over Poincare
recurrences, but argue that a proper understanding of how observable
cosmology emerges from the model, is consistent with the idea that
the model predicts the universe we see in a unique way. The
arguments of this paper are related to those of \cite{nightmare}
because they concentrate on the observations of localized observers,
rather than the overwhelmingly larger set of states on the
cosmological horizon.  In \cite{tbdS} I argued that there are many
horizon Hamiltonians which reproduce the same local physics, within
the maximum precision that local measurements can attain.} that this
finite system can undergo, there are many more histories which
contained very unlikely events (and thus {\it e.g.} have a current
CMB temperature so high that conventional extrapolation would
predict no surviving primordial nuclei besides hydrogen). Implicit
in this paradox is the claim that the time evolution we call
cosmology is the same as that defined by the static patch time in dS
space. In particular, different recurrences are supposed to
correspond to different cosmological histories. The claim of
\cite{DKS} is that, even if we restrict ourselves to histories in
which carbon based life could evolve, the typical history has more
entropy than our own.

\subsection{Io credo: a digression}

Any investigation of recurrences in this system has to deal with the
interpretation of the quantum theory of eternal inflation.   From
the global point of view, this has led to the problem of the proper
measure for counting bubbles of true and false vacuum.   The answers
one obtains depend crucially on the choice of measure, which is
surely determined by the correct theory of quantum gravity.   At the
present time, for cosmological space-times, the nature of the
correct theory is a matter of religious conjecture rather than
scientific or mathematical fact.   DKS implicitly used a measure
similar to that proposed (later) by Bousso {\it et.
al.}\cite{raphben}, and I will follow them.  I note that the problem
of Boltzmann's Brain, discussed below, may be resolved by other
choices of measure\cite{lindevil}.

In such a situation, one must begin by announcing one's own
religious credo:

\begin{itemize}

\item String theory provides us with many examples of consistent
quantum theories of space-time with asymptotically flat and low
curvature asymptotically AdS boundary conditions.   All of these
theories are exactly consistent with the ordinary laws of quantum
mechanics.

\item These theories provide evidence for the holographic principle.

\item A strong version of the holographic principle is that the
Hilbert space corresponding to a finite area causal diamond is
finite dimensional.   In particular, the quantum theory of a stable
dS space would have a finite number of states.   It immediately
follows that the mathematical quantum theory of such finite regions
is ambiguous: many mathematical quantum theories will give the same
results, within the precision allowed by the inevitable quantum
fluctuations in measuring devices constructed from a finite
dimensional quantum system.

\item Classical space-times with different asymptotics correspond to
quantum theories with different Hamiltonians. Therefore, different
classical solutions of the same low energy field equations may not
live in the same quantum theory.  In general, the Hamiltonian for
most space-times will be time dependent.

\item The correct quantum description of a theory of quantum gravity
may involve several different Hamiltonians.   Typically, some of
these are more fundamental, and the others are emergent descriptions
of a subset of degrees of freedom of the system, over time scales
where the fundamental Hamiltonians do not evolve those degrees of
freedom.

\end{itemize}

The first two of these principles are widely believed in the string
theory community.   The third was postulated by the author and W.
Fischler.   The covariant entropy bound refers to the entropy of
some maximally excited density matrix.   In a generic space-time,
with no asymptotic Killing vectors, there is no natural choice
besides the maximally uncertain density matrix, whose entropy counts
the logarithm of the total number of states.

The strongest evidence for the fourth principle comes from the
AdS/CFT correspondence, and from comparing its description of
Anti-de Sitter space, with the rather different descriptions of
asymptotically flat, and linear dilaton space-times.   More
generally, it follows from the fact that quantum theories are
differentiated by their high energy spectra, and that the spectrum
of black holes depends on the space-time asymptotics in a crucial
way.   Semi-classical quantization via the Wheeler-DeWitt equation
also leads to a time dependent Hamiltonian which depends on the
classical background.  Finally, if we accept the third principle, so
that dS space has a finite number of states, then the quantum
theories of dS spaces with different values of the c.c. are not
equivalent to each other.   Note however, that in this case, it
might make sense to think of a dS space with large c.c. as a
subsystem of one with small c.c. .   This is the premise of the
double well DKS model.

The last principle is exemplified by the description of in-falling
observers for black holes in asymptotically flat or AdS space-time.
Although the technical details remain to be worked out, a plausible
description is the following:   Consider, {\it e.g.} the
synchronous, Novikov, coordinate system, with initial time slice
coinciding with the $t = 0$ slice of Schwarzschild coordinates. The
quantum field theory Hilbert space on this slice, breaks up as a
tensor product of interior and exterior degrees of freedom.   We
restrict attention to initial states that do not lead to a
significant distortion of the background geometry (by conventional
semi-classical estimates).  This becomes a more and more restrictive
condition if we consider interior states, which become localized
near the singularity.   Thus, we should consider the Novikov
Hamiltonian an emergent Hamiltonian for a small subsystem, which is
only relevant for a limited period of time.

This should be contrasted with the Schwarzschild Hamiltonian, which
is an exact description of the system, as seen by asymptotic
observers\footnote{In the asymptotically flat case there is no exact
local time evolution, only a scattering matrix.}.   The two
Hamiltonians do not commute with each other.   In the local field
theory approximation, their failure to commute comes from regions
near the horizon of the black hole.   From the external point of
view, not much happens over Schwarzschild time periods of order the
in-fall time to the singularity.  The only interesting evolution
during this period is that experienced by the interior Novikov
degrees of freedom, under the Novikov Hamiltonian.  On time scales
much longer than this, the only valid description is that of the
external observer.   The interior thermalizes at the Hawking
temperature, and begins to radiate\footnote{We are imagining a black
hole formed from gravitational collapse, rather than an eternal
one.}.   There is no good description of the interior in terms of
field theory, and no meaning to the word {\it interior} in the exact
description\footnote{Rather, certain aspects of the interior
geometry can be read from asymptotic expansions of exact answers,
and have the limited meaning usual to asymptotic
expansions\cite{shenkerhong}.}. The interior Novikov Hamiltonian is
an emergent quantity, describing a subsystem for a limited period of
time.

\subsection{Un-digression}

I now want to argue that a similar situation exists in dS space.
Here the exact description of the system is associated with the
static dS Killing vector in a particular horizon volume.   The
emergent Hamiltonian is the time dependent Hamiltonian describing
cosmology, as well as the approximately Poincare invariant dynamics
of local physics, to which the cosmological Hamiltonian converges at
late times. I claim that neither of them is the static Hamiltonian
of dS space, and that even in the DKS model, the cosmological
evolution that is identified with the real world does not follow
from the dynamics of the static Hamiltonian. In fact, there are
strong indications that our cosmological time cannot be identified
with static dS time, except in the far future. In \cite{tbdS}, I
attempted to find a quantum theory that would account for all known
semi-classical results of quantum field theory in dS space. I
concluded that the system we call dS space must be described by two
Hamiltonians, acting in a Hilbert space whose dimension is
approximately the exponential of the Gibbons-Hawking entropy
$S_{GH}$. The static $H$ has a spectrum with spacing $c\ e^{-
S_{GH}}/ 2\pi R$.
 where
$c$ is a constant of order one, and $R$ the dS radius. $c T_{GH}$ is
the maximal eigenvalue of this Hamiltonian.

There are two arguments for this spectrum. One comes from an
examination of CDL tunneling rates and was mentioned in a footnote.
The other is that every excitation of dS space decays to the dS
vacuum. Since the vacuum shows no macroscopic evidence of even the
macroscopically large energies of black holes it is reasonable to
assume that black hole masses are not related to eigenvalues of the
static Hamiltonian.  Furthermore, the dS vacuum is an equilibrium
state with a unique temperature.   A generic state (with the flat
probability distribution on the unit sphere in Hilbert space) of a
random Hamiltonian obeying such a bound, approaches equilibrium at a
fixed temperature $T_{GH}$, as the number of states in the system is
taken to infinity. The constant $c$ is independent of the number of
states.   Fluctuations in the temperature in the ensemble of all
states in Hilbert space, are of order $e^{- {1\over 2} S_{GH}}$.
Thus, the assumption of a finite dimensional Hilbert space with a
Hamiltonian bounded by something of order the dS temperature,
explains the results of QFT in curved space for static dS space, as
well as CDL transition rates between dS minima.

This static Hamiltonian is the one that governs the long time scale
recurrences of dS space. In a static time $t \ll R$, a wave function
$| \psi > = \sum a_n | E_n > $ evolves very little.   Thus such a
Hamiltonian cannot describe the evolution of a universe which
changes significantly over time scales short compared to the Hubble
scale associated with the c.c..   It is also clear that its
eigenvalues can have nothing to do with ordinary particle masses or
energy levels of nuclei and atoms.    Instead, one proposes an
alternative Hamiltonian $P_0$, whose eigenvalues are related to
particle masses and other energies we measure.  $P_0$ is called the
Poincare Hamiltonian. Considerations based on the holographic
principle suggest a commutation relation, which is a finite
dimensional approximation to \eqn{cr}{ [H, P_0 ] = {1\over R} P_0 .}

Low lying eigenspaces of $P_0$ are thus approximately stable under
the $H$ time evolution.   In order that the thermal density matrix
$$\rho = Z^{-1} \ e^{ - H/T_{GH}}$$ look approximately like the
thermal density matrix

$$\rho_{EFT} = Z^{-1} e^{ - P_0 /T_{GH}} ,$$ one must postulate that
Poincare eigenvalues much less than $R M_P^2$ are equal to the
entropy deficit of the corresponding eigenspace, relative to the dS
vacuum. For eigenvalues $\gg M_P$, one can calculate the entropy
deficit from the black hole entropy formula, and it indeed agrees
with this expectation if we identify the black hole mass parameter
with the eigenvalue of $P_0$.

The operator $P_0$ removes the vast degeneracy of the spectrum of
$H$, for a small subset of states of the system. The Hilbert space
breaks up as

$${\cal H} = \bigoplus  {\cal H}_{p_0} \otimes {\cal K}_{p_0} ,$$
where $P_0$ is the difference between the GH entropy and the
logarithm of the dimension of ${\cal K}_{p_0}$.   The dimension of
${\cal H}_{p_0}$ is the degeneracy of the eigenvalue $p_0$ from the
point of view of ordinary physics. For black holes, this dimension
is the exponential of one quarter of the area of the black hole
horizon, while the entropy of ${\cal K}_{p_0}$ is that of the
cosmological horizon in the presence of the black hole.   The tensor
split of the Hilbert space described above is thus the split into
degrees of freedom localized near the observer, and those on its
cosmological horizon.  It is important to recall that the entropy in
localized excitations is bounded by something of order $(RM_P)^{3/2}
\ll (RM_P)^2 $, the entropy of dS space. $P_0$ encodes the
information relevant for local measurements.   It is useful on
time-scales short compared to the decay time of localized
excitations.   A commutation relation between $H$ and $P_0$ roughly
like that given in \ref{cr} follows from this description of the
spectra of the two operators.

On long enough time scales, where time is defined in terms of the
evolution under $H$, none of the eigenstates of $P_0$ is stable. For
large $p_0$  the decay time is of order $p_0^3 l_P^4 $.   For some
of the lower lying eigenstates it is exponential in the dS radius
\footnote{This is the time one has to wait for a thermal fluctuation
to destroy a localized object which is not a black hole.}.   Once
localized systems decay, and their decay products disappear through
the horizon \footnote{The phrase {\it decay products} includes
galaxies that are not gravitationally bound to us.} the dynamics
described by evolution under $P_0$ is no longer relevant {\it for
those degrees of freedom}. Eventually, all localized excitations
disappear (until next time) and the very slow time evolution
generated by $H$ is the asymptotic dynamics of the system.

The Hamiltonian $P_0$ is thus an emergent feature of the dynamics of
de Sitter space.   Once a localized object has appeared, through a
fluctuation, the dynamics of $H$ does not change it at all, over
time scales $ < R$, and in some cases much longer time scales,
though still short compared to the recurrence time $e^{\pi
(RM_P)^2}$. For these short times the Hamiltonian $P_0$ evolves the
localized degrees of freedom, by resolving the huge degeneracy of
$H$. The Hamiltonian $P_0$ was introduced in order to understand how
a Poincare invariant theory describing particle physics could emerge
from the theory of dS space in the large $R$ limit. In a cosmology
which is only future asymptotically dS, the analog of $P_0$ is a
time dependent $P_0 (t)$.   For the semi-classical cosmology of DKS,
one would compute it approximately by studying quantum field theory
in the future Lorentzian CDL bubble, with boundary conditions
determined by analytic continuation from the Euclidean section.
There are important corrections to the classical background during
the period of reheating, when oscillations of the scalar field can
decay into radiation.

What is important however is that once we have established the
correct background classical geometry, we must restrict attention to
states of the QFT which do not have significant back-reaction on the
geometry. The Hilbert space of degrees of freedom on which $P_0 (t)$
acts, consists only of those small fluctuations.   Other classical
homogeneous background geometries, which do not evolve from the CDL
initial conditions are not part of the quantum theory, according to
religious principle number $4$ of the previous subsection.

The time dependent Hamiltonian, $P_0 (t)$, of this system will, for
this subset of states, approach the Poincare Hamiltonian $P_0$ as
cosmological time gets large.   On a longer time scale - that of the
disappearance of all localized degrees of freedom through
catastrophic thermal fluctuations followed by Hubble flow of the
debris out through the horizon - the slow dynamics of the static
Hamiltonian $H$ becomes relevant.

The Poincare recurrences discussed in \cite{DKS}, are a feature of
the $H$ dynamics.   The Hamiltonian $P_0$ and its time dependent
cosmological cousin, do not have such recurrences.   Indeed they are
only relevant over time scales at most of order $e^{c (R
M_P)^{3/2}}$, with $c$ a constant of order $1$.   Let us try to see
whether we can reproduce the paradox of \cite{DKS} with our new
understanding of the dynamics.   Most of the time (H evolution) our
system resembles the dead dS vacuum.   Interesting observers can
only exist in the period following a low entropy fluctuation.   One
such fluctuation is the CDL tunneling which produces a system which
resembles a dS vacuum state with c.c. $\Lambda_1$ .   Inflation
rapidly wipes out any non-vacuum excitations of this state, and the
reverse tunneling event automatically produces an approximately
homogeneous and isotropic FRW universe.   For a long time on real
cosmological scales, the evolution according to the Hamiltonian $H$
is irrelevant, because the bound on the spectrum of $H$ tells us
that the wave function has not changed.   {\it Instead, the dynamics
of cosmological observers is governed by $P_0 (t)$. It is extremely
important to note that all of conventional cosmology takes place
during this period.  The dynamics described by $H$ becomes relevant
only when all of the localized excitations produced by the tunneling
event, have decayed or exited through the cosmological
horizon}\footnote{This is an exaggeration.   Some localized
excitations decay only because an unlikely thermal fluctuation
destroys them.   In principle the microscopic description of this
fluctuation is governed by the $H$ dynamics, though for all
practical purposes it may be that it can be described by the thermal
physics of $P_0$. }.

One should remember that the dynamics described by $P_0 (t)$ only
involves a small subset of the degrees of freedom of the system. The
entropy of the relevant localized degrees of freedom is bounded by
$c (RM_P)^{3/2}$.  For the rest of the degrees of freedom, only the
time evolution defined by $H$ is relevant.  Not much happens to them
on the time scale defined by cosmology.   However, as things begin
to go out of the cosmological horizon of the large dS space, fewer
and fewer degrees of freedom of the original cosmological
fluctuation participate in the $P_0 (t) \rightarrow P_0$ dynamics.
Thus, the time evolution which describes cosmology is an emergent
property of the semiclassical low entropy fluctuation.   Eventually,
on time scales $\gg R$, when we return to the equilibrium dS vacuum,
there are no degrees of freedom to which to apply the $P_0 (t)$
Hamiltonian and only the $H$ evolution is relevant.

Future recurrences of the events which led to semi-classical
cosmology will not lead a substantially different history.   First
of all, by the assumption that the process went through tunneling,
the entropy of possible initial states is just that of the small
radius de Sitter space.   Secondly, traces of the initial state will
be wiped out by false vacuum inflation. Finally the slow roll
inflation in the basin of attraction of the true minimum will, for
appropriate values of $\mu$, smooth out initial inhomogeneities and
leave only the slow roll inflationary fluctuations to be seen in the
CMB.   For a small number of e-foldings, such as we expect from
field theoretically natural potentials, there may be evidence of the
initial conditions left at the largest cosmological scales.

There is no reason to suspect that recurrences can lead to
situations like a cosmology with higher CMB temperature, as
envisioned by DKS.   Cosmology is time evolution under the emergent
Hamiltonian $P_0 (t)$, which is {\it defined} in terms of a
particular semi-classical fluctuation of the dS background.   There
are no CDL instantons which describe an alternative cosmological
history differing only in the temperature of the CMB.   The dominant
quantum/thermal fluctuations around the dS background produced
localized excitations, and we have already seen that localized
fluctuations around the CDL background do not have a significant
effect on cosmology.   If we try to extrapolate {\it generic} late
time dS fluctuations back to the past, we find a Big Bang, rather
than the CDL cosmology\cite{landskeptbf} .   This cosmology is not
only out of semi-classical control, but unstable. If we make small
changes in the initial conditions near the Big Bang, we will produce
a universe which does not have an inhabitable future.  There is of
course no argument that actual recurrences in the time evolution
generated by $H$ will actually lead to any of these semi-classical
cosmologies.

To summarize, in a system evolving under the dS Hamiltonian $H$, if
the potential characterizing the low energy degrees of freedom has a
subsidiary minimum with larger c.c., then the CDL instanton gives us
evidence for transition to a state which looks like an evolving
cosmology.   The time evolution in this cosmology is {\it not} that
of $H$, but of an emergent time dependent Hamiltonian $P_0 (t)$,
which gives an approximate description of a small subset of
localized degrees of freedom.   The time scales for $P_0 (t)$
include some which are much shorter than $R$ and none which are
larger than $e^{c (RM_P)^{3/2}}$\footnote{This is the time scale
over which localized measurements become impossible.   No localized
device is free from quantum fluctuations of its pointers on any
longer time scale. In reality, it is very improbable for any
localized system to survive a time longer than $e^{2\pi MR}$, where
$M$ is its Poincare eigenvalue. Within a few times $R$, the
evolution of $P_0 (t)$ is applicable only to the small subset of the
(currently) visible universe that is gravitationally bound to the
observer. If this system has mass $M$, it can collapse into a black
hole, and decay to the dS vacuum in a time of order ${M^3 \over
M_P^6}$, or survive until a thermal dS fluctuation annihilates it in
a time of order $e^{2\pi RM}$.   In any case, {\bf LONG} before the
recurrence time, $e^{\pi (RM_P)^2 }$, the cosmological Hamiltonian
has \lq\lq de-emerged ".   There are no longer any degrees of
freedom to which this time evolution applies. }.

Since the $P_0 (t)$ evolution that describes cosmology is completely
decoupled from the recurrences, we no longer have any plausible
evidence, let alone a compelling argument, that the system has the
kind of anthropically allowed, but observationally forbidden
histories that DKS imagined\footnote{There is one kind of recurrence
of anthropically allowed, but observationally forbidden histories
which is not ruled out by this argument.   We will deal with it
below in the section entitled {\it Boltzmann's Brain}. }.   Every
recurrence of cosmological history that we can study semiclassically
resembles every other in all but the finest details.   It is
possible that we can actually observe some of those details in the
future, but virtually certain that we have not yet done so.

Effective field theorists will object that there are other solutions
of the low energy field equations, which describe the misanthropic
cosmologies of DKS.   Although singularities occur in the backward
extrapolation of most modifications of the CDL geometry, there are
plenty of other homogeneous isotropic solutions of the field
equations. I have argued long and hard and to no avail\cite{tbargue}
that different solutions of low energy effective field equations are
not necessarily part of the same quantum theory of gravity.  This
was religious principle $4$ of the previous subsection.

In the present context I believe we have strong evidence that small
quantum fluctuations around the dS and CDL backgrounds, and black
hole states in the dS background are really excitations of the
quantum Hamiltonian $H$. Note however that most of the black hole
solutions are not allowed during the cosmological evolution in the
CDL background. Only those black holes which form as a consequence
of evolution from the non-singular CDL background will appear. Large
black holes that do not form as a consequence of cosmological
evolution, can only appear as thermal fluctuations.  This will occur
after the normal cosmological history has ended, and the system
returns to its dS equilibrium. Thus, our system really has 3
different interesting Hamiltonians, $H$, $P_0$, and $P_0 (t)$. $P_0$
describes the Lorentz invariant particle physics which emerges in
the $R \rightarrow\infty$ limit, and finite $R$ corrections to it
like SUSY breaking\cite{tbfolly}. $P_0 (t)$ describes the DKS
cosmology. At late times, low eigenvalues of $P_0$ and $P_0 (t)$ (in
the sense of the adiabatic theorem) coincide, but $P_0 (t)$ has no
analog of the large black hole eigenstates of $P_0$.   $H$ describes
the long time evolution of the system and is the only full
description of it.

My claim is that we have no reason to assume that other classical
solutions of the low energy field equations have anything to do with
our quantum system, until someone shows us a calculation at least as
compelling as that of CDL for computing transition probabilities
between empty dS space and these other semiclassical geometries.
These transition probabilities should satisfy a law of detailed
balance compatible with the entropic characteristics of the system.

The idea that evolution operators in quantum gravity, are tailored
to a specific classical background, is familiar from the resolution
of the Problem of Time\cite{tbrubetalbfs} in the Wheeler-DeWitt
approach.   The solution space of the hyperbolic Wheeler-DeWitt
equation does not possess a positive definite metric.   Expansion
around a specific classical solution reduces the equation to a
parabolic time dependent Schrodinger equation, which does. Thus,
different classical solutions of Einstein's equations lead to
different quantum theories.   Some of these may be the same theory
as viewed by different classical observers, but there is no reason
to expect that to be a general feature, since it is not even true in
the semi-classical approximation. What is peculiar in the present
context is that DKS thought they were dealing with such a fixed time
evolution, under the static Hamiltonian $H$. However, the
characteristics of the spectrum of $H$, which are necessary to
explain the global semi-classical properties of the causal patch of
dS space, imply that it cannot be identified with ordinary energy,
or with the cosmological evolution operator $P_0 (t)$.   The
observers that experience cosmological evolution, as described by
$P_0 (t)$ are built out of localized degrees of freedom.   They
cannot experience recurrences.

We have also argued that every recurrent appearance of this sort of
localized observer will evolve in practically the same manner.   The
DKS cosmology thus provides a rationale for a low entropy beginning
of the universe as a fluctuation. As a bonus, it suggests that the
entire history of the universe can be described by semi-classical
physics. There is no singular Big Bang.

\subsection{Landscapism}

So far, the considerations of this paper apply to all potential
landscapes with potentials {\it above the great divide}\cite{abj} .
The CDL transition probabilities out of the dS minimum with smallest
c.c. are all suppressed by an entropy factor, and it is consistent
to postulate a system with a finite number of states.   It should be
noted that in such a system it is possible for anthropic
considerations to trump purely dynamical expectations for how long
the system spends in each of its meta-stable states.   The authors
of \cite{klebanetal} have emphasized that in systems of this type,
galaxy formation requires {\it both} a small c.c. and a period of
slow roll inflation.  Thus, although CDL probabilities predict that
the system will spend the vast bulk of its time in the basin of
attraction of the minimum with lowest positive c.c., this may not be
a region in which galaxies can form.   Most galaxies will be found
in cosmologies based on a CDL instanton whose Lorentzian section
lives in the basin of attraction with smallest c.c., subject to the
constraint that there be enough e-folds of inflation\footnote{The
real situation is of course much more complicated.   Different
minima will have different low energy degrees of freedom.   It is
possible that some minima will not have appropriate matter from
which to form galaxies.}.

Unfortunately, we cannot apply these considerations to the only
landscape that has any sort of theoretical underpinning, The
Landscape of String Theory.   The string theory landscape certainly
contains {\it terminal vacua}\footnote{To use
Boussonesque\cite{bousso} language, if not his equations.}, which
are asymptotically SUSic and do not allow for recurrences.
Furthermore, it is likely that the string theory landscape lies
below the great divide: that is, it is replete with
non-supersymmetric, small c.c. dS minima which tunnel rapidly to a
Big Crunch. This is also inconsistent with a model having a finite
entropy.

Thus, although the model of DKS can be analyzed in a predominantly
semi-classical manner, and, within the semi-classical approximation,
provides a real solution to Penrose's puzzle about initial
conditions, it is not clear that it is a semi-classical
approximation to a real model of quantum gravity.

\section{Boltzmann's brain}

There is one kind of ``anthropically allowed" but observationally
forbidden history that is not ruled out by our previous
considerations.  This is the situation colloquially known as {\it
Boltzmann's brain}.    Our system cycles through all of its allowed
states, populating them statistically.   Suppose that we consider a
state which looks like the true vacuum dS space, marred only by a
fluctuation consisting of Boltzmann's brain, replete with all of its
memories.   The probability for this can be estimated by computing
the probability for nucleating a black hole with the same mass as
the brain, and then dividing by the number of localized states of
the black hole.   This gives\cite{gpbh}

$$P_{brain} \simeq e^{- 2\pi m_b R} e^{- \pi ({m_b \over M_P})^2}
.$$   The probability is small, but much larger than the CDL
probability to produce the DKS cosmology.    The claim is that this
implies that the dominant form of intelligent life in the DKS model
is a brain spontaneously nucleated in empty dS space with all of its
memories of the CDL instanton cosmology.  Of course, such a brain
would immediately explode and/or die of oxygen starvation but we can
solve this by nucleating a support system for the brain, still with
exponentially higher probability than the CDL process.   The real
prediction is that the dominant form of intelligent life in the DKS
universe is a form created spontaneously with knowledge of a
spurious history, which lives just long enough to realize that its
memories are faulty.

Pursuit of this line of reasoning leads us into two opposite
directions, both somewhat speculative.   The first is the
theoretical bound on the mass and radius of the smallest intelligent
system that will ever be constructed in our cosmological
history\footnote{Like all anthropic questions, we must narrow
ourselves to local physics and chemistry like our own in order to do
any calculation at all.   We will probably never have the
theoretical power to determine whether intelligent life would have
been possible with other low energy gauge groups and representation
content.}.   This definitely includes artificial intelligences that
our descendants might someday create.   Certainly, unless we declare
{\it a priori}  that silicon based machines will never have human
scale intelligence, the smallest viable version of Boltzmann's brain
will be many orders of magnitude smaller and lighter than
Boltzmann's own, as well as much more robust.   R.
Bousso\footnote{Private communication, later repudiated.} has
suggested $m_b r_b > 10^{25}$ as a bound on the minimum mass and
size of a brain.

The other direction to take is to speculate on the possibility that
the minimal ``support system" for a brain is actually very large.
Our immediate reaction to the proposal for spontaneous nucleation of
Boltzmann's brain is that it is ridiculous.   It is ridiculous
because we know about the complex and tortuous history that went in
to making the original version.   We had to make galaxies and the
galaxies had to make the right kind of stars and planets and there
had to be a planet at the right distance from the Sun and it had to
undergo both long periods of relative stasis as well as short
catastrophic periods of dramatic environmental change.   And even
then we don't know how much of an accident intelligent life was,
because we really only have one data point.

In fact, I believe this intuitive reaction has some validity.   What
we call {\it the state of the brain} or {\it the state of a galaxy}
in our universe is in fact not a quantum state of those subsystems
of the universe, but an entangled state of those subsystems with
other degrees of freedom in the universe\footnote{Lest the reader
think that I am advocating some kind of weird New Age effect of
quantum mechanics on consciousness, I would ask her/him to recall
{\it e.g.} that the classical force between static sources is, in
quantum field theory, encoded in just such a phase correlation in
the wave function describing the two sources.}.   In this sense, the
past is encoded in the current state of Boltzmann's brain, which
cannot be properly described in terms of any wave function for the
degrees of freedom in the brain itself.

One can imagine constructing a mathematical argument which would
turn this observation into a refutation of the claim that
Boltzmann's brain and its memories can be created by a thermal
fluctuation, because the brain would not function without the phase
correlations with the rest of the universe that encode its past. I
have not yet done so. It is likely that even this appeal to past
history might not be sufficient to resolve the problem.   The most
important quantum correlations between different parts of the
universe are those encoded in Newtonian forces. A. Aguirre has
pointed out to me that in simulations of galaxy formation, it seems
sufficient to restrict attention to a region of size $(10\ {\rm
megaparsecs})^3$ surrounding our galaxy (Boltzmann's local group).
That is, a region of this size will evolve like our own surroundings
did, for 10 - 15 billion years, even if it is embedded in flat
space. The rest of the universe doesn't seem to matter. The thermal
probability for producing such a region in precisely its current
state, is small, but still larger than the CDL probability, for
producing the full history of the DKS cosmology as a fluctuation.
Thus, while the records of past history encoded in the actual
correlations between Boltzmann's brain and the rest of the universe
(some of which we feel in particular as Newtonian forces) might help
to explain our conundrum, we have to do more work to show that the
most probable way to obtain our local group of galaxies as a
fluctuation in the DKS model is to follow the full course of CDL
evolution.    These observations might resolve the string landscape
version of the Boltzmann's brain paradox, to which we now turn.

\subsection{Boltzmann's brain in the string landscape}

Susskind has emphasized that one of the attractive features of the
string landscape is that it seems to avoid the problems of the DKS
model, because it does not have general recurrences.   That is, in
the string landscape, meta-stable dS vacua decay into terminal vacua
where no life can exist, on a time scale no longer than the
recurrence time. Thus the kind of misanthropic cosmological
histories envisaged by DKS cannot occur in the landscape.

However, we have seen that cosmologies with a slightly higher CMB
temperature are not really a problem in the DKS model.   The real
problem is Boltzmann's brain.    If we ignore the correlations
between the state of the brain and the rest of the universe, which
we alluded to in the last paragraphs of the previous subsection,
then Boltzmann's brain might be a problem for the landscape.   Using
Bousso's estimates for the minimal size and weight of any brain, the
probability of thermal fluctuations producing a brain is $e^{-
10^{25}} $.   CDL decay probabilities for moduli fields with
potentials below the Great Divide are of order $ e^{ - ({m_P \over
\mu})^4}$, where $\mu$ is the typical scale of the potential. These
will be larger than the brain production probability only if $\mu >
10^{-6} m_P $.   Note that, in order to produce thermal brains, we
only need one vacuum with low energy physics like our own, and
cosmological horizon size $> 1$ cm. (Bousso's estimate of the
minimal size of a reasonable brain), which violates this bound on
$\mu$ (or the analogous bound for Brown-Teitelboim tunneling).

At the level of validity of all arguments about the String
Landscape, it is fairly certain\cite{dinelindeshenkeretal}, that
there are approximately SUSic dS vacua in the landscape, which lie
above the great divide.  These have all decay probabilities of order
$e^{- \pi (RM_P)^2}$, much smaller than the probabilities for
producing BBs .   Thus, in order for the landscape to completely
evade the problem of BBs, one must establish that none of these
nearly SUSic vacua have low energy physics resembling our own. This
seems fairly unlikely, given the reasonable success that string
theorists have had in finding supersymmetric states with more or
less the standard model in various sorts of perturbative string
vacua.   Thus, it seems rather unlikely that the landscape will
completely avoid confronting recurrences of the great Austrian
physicist.

In the landscape, in contrast to the DKS model, it is harder to
estimate the probability of producing complex intelligence by the
historically favored route, so even if the landscape produces
Boltzmann brains, we don't know with certainty that they are the
most probable form of intelligence. This is a direct consequence of
the fact that the landscape {\it does not deal with the problem of
low entropy initial conditions at all.} It is reputed to have an
infinite number of quantum states, so the Boltzmann/DKS argument
does not apply. Discussions of cosmology in the landscape generally
begin with a CDL tunneling event, but no explanation is given of why
the universe ever got into the low entropy state from which this
tunneling proceeds.

In the next section we will explore a completely different mode of
explaining low entropy initial conditions.

\section{Holographic cosmology}

\subsection{The roots of holographic cosmology}

Holographic cosmology does not fit into the paradigm of quantum
gravity as a Feynman path integral over space time metrics.   It
imposes a fixed causal structure and a fixed foliation on space-time
as a matter of principle.   This is because it focuses on the
physics that can in principle be measured by a single observer. The
\lq\lq relativity " of general relativity is enforced by considering
a whole collection of observers, with partially overlapping degrees
of freedom, and insisting that they all give a consistent
description of the physics in the overlapping regions.

In quantum mechanics, an observer is a large system with many
semiclassical degrees of freedom that can serve as pointer
observables.   Our mathematical understanding of such systems is
based on local field theory (perhaps with a cutoff), and the
pointers are averages of local fields over volumes much larger than
the cutoff scale.   Ideal quantum measurements refer to the limit of
infinitely large pointers. The existence of such observers in
localized regions of space-time is at odds with the holographic
principle, and with our understanding of semi-classical general
relativity.   In GR we expect that if we put too much mass into a
region, we form a black hole.   The holographic principle formalizes
this expectation as a bound on the entropy of certain space time
regions.   Consider in particular a causal diamond, the intersection
of the interior of the backward light cone of a point P with the
interior of the future light cone of a point Q in the causal past of
P.   We can think of these as two points along the worldline of a
time-like observer, and we can build up the world line itself in
terms of a nested sequence of causal diamonds.   In classical GR,
the causal diamond represents the entire region on which an observer
traveling from Q to P can do experiments.

The covariant entropy bound\cite{fsb} is a conjectured bound on the
entropy of the causal diamond, by (one fourth of) the area in Planck
units of the maximal area spacelike surface on the boundary of the
diamond.   For sufficiently small time-like interval between P and
Q, this is always finite.   The bound can be proved from Einstein's
equations with additional assumptions bounding entropy density by
energy density\cite{marolf}.   However, it is clearly a statement
about quantum gravity and cannot be proven without a theory of
quantum gravity.   In this respect, string theory is not much help.
String theory does not talk about local physics.   In asymptotically
flat space-time the only causal diamond that appears in string
theory is the conformal boundary.   In asymptotically AdS space-time
one can give unambiguous meaning only to the causal diamonds of
points P and Q whose time-like separation is large enough for the
causal diamond to intersect the boundary\footnote{More localized
measurements in AdS/CFT are related to ambiguous choices of UV
cutoffs on the boundary field theory.}.

Our idea has been to take the holographic principle and the entropy
bound as the defining property of quantum gravity.   To do this, one
has to decide which quantum density matrix the bound refers to.
Fischler and I argued that the only general answer could be the
maximally uncertain one: {\it the covariant entropy bound for a
causal diamond refers to the logarithm of the dimension of the
Hilbert space which describes experiments that can be performed by
the observer traveling between Q and P.}   Other, more restrictive
density matrices would only be natural if there were a preferred
operator, like the Hamiltonian, in our system.  The message of
general covariance is that there should be no such preferred
operators except for infinite systems.   Energy and other symmetry
generators are only defined in terms of asymptotic diffeomorphisms
acting on the boundary of an infinite space-time.

The idea that the theory of a finite causal diamond has a finite
number of states, immediately implies that that theory is ambiguous.
A quantum theory with a finite number of states cannot perform
measurements on itself with arbitrary accuracy.   If the {\it a
priori} quantum uncertainty in the results of measurements is
irreducibly finite, then many different mathematical quantum
theories will predict the same answers within the unavoidable
measurement error.   This is what I mean by the statement that {\it
the quantum theory of a finite area causal diamond is intrinsically
ambiguous.} We call a choice of a particular mathematical quantum
theory for a sequence of causal diamonds a {\it choice of a quantum
observer}. For asymptotically large sequences of causal diamonds,
with the property that the asymptotic limit is well described by
quantum field theory in curved space-time (this is a restriction on
the allowed asymptotic states of the system), this notion approaches
the notion of observer in classical GR and ordinary quantum theory.
That is, under such conditions, we know that there will exist large
quantum subsystems with many semi-classical observables.   In this
limit, we can easily compare the results obtained by different
causally connected observers\footnote{The proper definition of what
we mean by different observers will be given in a moment.} because
their semi-classical observables commute with each other.   The
usual equivalence relations imposed by GR should apply, but only in
the limit.

\subsection{Executive summary}

Before proceeding to a long discussion of holographic cosmology, I
want to summarize the results I intend to demonstrate. This should
be useful for the small number of readers who have followed the
details of previous papers on holographic cosmology, and the much
larger group, which has no interest in those details.

Holographic cosmology is an attempt to study generic initial
conditions at a Big Bang singularity in terms of quantum mechanical
models built to satisfy the holographic principle.   The claim is
that generic initial conditions lead to a state which can be
colloquially described by saying that the particle horizon of each
observer is, at all times, filled with a maximal size black hole.
Averaging over many horizon volumes, this defines a homogeneous,
isotropic, flat universe, with equation of state $p = \rho$. This
system is called the {\it dense black hole fluid}. No real observers
can exist in such a universe, because all of its degrees of freedom
are always in equilibrium.

The real universe is supposed to arise as a lower entropy initial
configuration, which has maximal entropy among all those which avoid
collapse into a dense black hole fluid state.   We argued that these
initial conditions make a transition to a nearly homogeneous dilute
black hole gas at a critical value of the particle horizon size,
$M$.  The inhomogeneous fluctuations are exactly scale invariant
over a range of scales whose physical size ranges from the Planck
length to $M$ at the time of the transition.   Their amplitude is
small, but neither its exact value, nor the value for $M$, can be
calculated at present.

These initial conditions are such that inflation is relatively
probable if the low energy effective field theory has a field with a
relatively flat potential.   Only around $20$ e-folds of inflation
are needed to explain the correlations in the CMB, in a causal
manner. The observable fluctuations in the CMB could be generated
either in the $p = \rho$ era or during the inflationary era.   If
experimental indications that the fluctuations are not exactly scale
invariant hold up, then the CMB fluctuations must come from
inflation.   The energy scale during inflation in holographic
cosmology is less than $10^{-3} M^{-2}$ in Planck units, and $M$
itself is likely to be $>> 1$. Inflationary fluctuations in such a
low scale model can be consistent with the data only in certain
kinds of hybrid inflation models.   Such models usually require
either fine tuning or supersymmetry.     Also, since the scale of
inflation is low, the ratio of tensor to scalar modes is small.

We have not been able to find a significant source of primordial
gravitational waves in the pre-inflationary physics of the
holographic model, so one would tentatively conclude that
holographic cosmology predicts no observable primordial
gravitational waves.

The final state of a holographic cosmology is a stable de Sitter
space.   Since the normal region, which evolves to this dS space is
a low entropy fluctuation of the $p = \rho$ fluid, purely
statistical arguments favor the largest value of the cosmological
constant.  This requires the smallest deviation from the maximal
entropy configuration.   However, if we want the normal region to
contain galaxies, the cosmological constant is bounded from
above\cite{wein}.   The connection between the value of the c.c. and
the scale of SUSY breaking\cite{tbfolly} may provide an even
stronger upper bound\cite{penta}.  The current model\cite{penta} of
low energy physics based on these ideas also gives a lower bound on
$\Lambda$.  Thus it is at least possible that we will end up
predicting that the value of the c.c. is determined by a combination
of statistical arguments and the requirement that certain gross
features of low energy physics are reproduced by our model.

Now turn to the problem of Boltzmann's brain in holographic
cosmology. The stable dS endpoint of holographic cosmology will
certainly produce Boltzmann brains, if such objects can exist.
However, in contrast to the DKS model, ordinary cosmology is not an
extremely low entropy fluctuation of our system.   The system has a
fixed initial condition (the Big Bang) and we will try to argue that
a cosmology like our own is the most probable result of initial
conditions which do not produce a dense black hole fluid.  This
argument depends on the claim that there is a theory of stable dS
space for every value of the c.c. $\ll$ the Planck scale, and that
the limiting theory for vanishing c.c. is exactly supersymmetric and
has a compact moduli space. Since there are no examples of such
isolated SUSic theories in which we can do reliable calculations
(the state of the art is \cite{dsbbv}), it is safe to say that the
possible theories of stable dS space are much less numerous than the
points in the hypothetical String Landscape.   It is even reasonable
to assume that there is only a unique such theory, which one would
of course hypothesize to have low energy physics coinciding with
that in the real world. If this is the case, then production of
intelligent life by evolution is a natural consequence of our
model\footnote{and might one day be proved to be a consequence of
the model with high probability.}.  By contrast, Boltzmann brains
are low probability events, which produce short lived intelligence.
It is difficult, though perhaps not impossible, to imagine
constructing devices during the evolutionary era of the model (which
we identify with our own), which have high enough resolution to
detect small Boltzmann brains, but whose recording devices have
quantum fluctuations sufficiently small to be ignored for times of
order the waiting time to produce a Boltzmann brain. Thus, in
holographic cosmology, Boltzmann brains are in principle observable
by the much more probable intelligent organisms (our descendants?)
produced by the inevitable cosmological evolution of the model. They
are freaks of nature and pose no philosophical conundrums.

The issue of {\it Poincare Recurrences} has also been raised, as a
criticism of models which, like holographic cosmology, terminate in
an asymptotically de Sitter system, with a finite number of
states\cite{DKS}.  In such systems, the asymptotic de Sitter
Hamiltonian has Poincare recurrences.  In holographic cosmology, the
universe we observe, over the time that we currently characterize as
its entire history, is described by an emergent time dependent
Hamiltonian. The entropy of such states scales like $(RM_P)^{3/2}$
as the dS radius goes to infinity. As noted above, the fundamental
claim of holographic cosmology is that normal semiclassical
cosmological evolution, is a high probability result of those
initial conditions that do not produce horizon filling black holes
at every cosmological time. In this view, normal cosmological
evolution is not described by the asymptotic dS Hamiltonian, and
that {\it history} will never recur.   Poincare recurrences may
produce some of the states encountered along this history, but will
not evolve them in the same way that the time dependent Hamiltonian
does.   Thus, in this model, Poincare recurrences will {\it never}
reproduce the cosmological evolution of intelligence that we believe
occurred in the world we observe.

Finally, in the case of Poincare recurrences, general arguments
\cite{nightmare} show that no machine built out of local quantum
field theory degrees of freedom\footnote{Unless one can build
complicated machines using the classical degrees of freedom of black
holes, in contradiction to the No Hair theorem, this means no
machines that can exist in any quantum theory of gravity.} can
remain classical long enough to observe a Poincare recurrence. This
means that, unlike Boltzmann brains, there is {\it in principle} no
operational way to test for the existence of Poincare recurrences.
Mathematically, this means that we can build many models of the
universe of holographic cosmology, all of which make the same
predictions for observations about local physics and the
evolutionary part of cosmic history, but which differ in their
predictions for the behavior of the system over a Poincare
recurrence time.   It seems most reasonable to declare all such
descriptions to be gauge equivalent, and the mathematics of Poincare
recurrences to refer to gauge degrees of freedom.

\subsection{Holographic cosmology: the details}

We want to construct a microscopic theory, which is purely quantum
mechanical, and mimics the causal structure and equivalence
relations for observers in general relativity or quantum field
theory in curved space time.    We define an observer by a nested
sequence of Hilbert spaces ${\cal H} (t) = {\cal K} \otimes {\cal H}
(t - 1) $.   These represent the degrees of freedom accessible to
the observer in a nested sequence of causal diamonds, which
eventually engulf its entire history. In the cosmological context,
the only one we will discuss in this paper, we should think of these
spaces as associated with a nested sequence of causal diamonds, with
their past tips lying on the same point on the Big Bang surface. The
growth of the Hilbert space with $t$ refers to the growth of the
particle horizon of the observer, as its trajectory gets further
from the Big Bang. The finite dimensional Hilbert space ${\cal K}$,
refers to the degrees of freedom in a single pixel of the observer's
holographic screen, and we will discuss its structure in a moment.
As the particle horizon grows, more pixels are added to the screen.

The dynamics in each of these Hilbert spaces is given by a sequence
of unitary operators $U(t,k) $ with $1 \leq k \leq t$, obeying the
consistency conditions
$$U(t, k) = U(s,k) \otimes V(t,s,k) ,$$ whenever $t \geq s \geq k$.
As the tensor product indicates, $V(s,t,k)$ operates only on that
tensor factor of the Hilbert space ${\cal H} (t)$, which is
complementary to ${\cal H} (s)$.  In words: as the particle horizon
expands the observer sees new degrees of freedom, which couple to
those already within its purview.  The consistency condition
guarantees that the past as seen by an observer at a future time is
not changed from that seen by the earlier observers.

A full quantum cosmology is defined by a spatial lattice of such
observers.   Only the topology of the lattice has meaning. Geometry
of the space-time is supposed to emerge as a coarse grained
approximation in the limit of large causal diamonds.  Together with
the time lattice of each observer, the space-lattice defines a
discrete topological space-time.   It has a causal structure, and a
global foliation by space-like surfaces.   The time function, is a
monotonically increasing function of the area of the causal diamond
whose future tip lies on each space-like surface.  The full causal
structure is defined by specifying, for each pair of Hilbert spaces
${\cal H} (t, {\bf x} ) $ and ${\cal H} (s, {\bf y})$ an overlap
Hilbert space  ${\cal O} (t, s, {\bf x,y})$, which is a tensor
factor of each.   The overlaps are constrained so that one gets the
same result for ${\cal O} (t, s, {\bf x,y})$ by moving along any
path between the points.   Furthermore, for nearest neighbor spatial
lattice points  ${\cal O} (t,t, {\bf x, x + \hat{n}}) = {\cal K}$.
Finally, the dimension of ${\cal O} (t,t, {\bf x,y})$ should
decrease monotonically with the inverse length of the shortest
lattice path between ${\bf x}$ and ${\bf y}$.   Two points at fixed
time are causally disconnected when ${\cal O}$ is one dimensional.

On the overlap Hilbert space, we must also insist that the unitary
evolution operators defined by the individual observers agree with
each other\footnote{Perhaps only up to unitary conjugation: $ U(t,
k, {\bf x} ) = W U(t  , k , {\bf y}) W^{\dagger} $ on ${\cal O}(t,t,
{\bf x,y})$.   The model we will study has $W = 1$.}  This is an
incredibly complicated condition, and we think of it as the
dynamical principle of holographic space-time.   Different solutions
to it will be the allowed quantum cosmologies of this formalism.

Here is the only solution to it that we have found.   Let ${\cal O}
(t,t, {\bf x, y}) = {\cal H} (t - N, {\bf x}) $ where $N$ is the
number of steps in the shortest lattice path between ${\bf x}$ and
${\bf y}$.   Other overlaps are computed by combining this rule with
the rule for overlaps along the trajectory of a given observer.
Since ${\cal H} (t, {\bf x})$ is the same Hilbert space for every
${\bf x}$, this rule can give consistent dynamics if (and only if)
we choose the same sequence of unitary operators for the observer at
each ${\bf x}$.   Thus, our formalism for quantum cosmology gives a
consistent result in this case only if the cosmology is spatially
homogeneous.   Similarly, for large $k$, the locus of points that
are $k$ lattice steps away from a given point fills out a sphere in
the Euclidean space in which the lattice is embedded, as long as the
lattice has the topology of Euclidean space.   Thus for large $t$,
the distance at which two observers become causally disconnected
becomes spherically symmetric.   Our cosmology is homogeneous,
isotropic, and flat.

To understand the rest of the dynamics we must discuss a point we
have omitted, the nature of the quantum variables which describe
cosmology.   Consider the holographic screen of a causal diamond. It
is a space-like surface, which records information from the interior
of the diamond when a massless particle penetrates it.   A {\it
pixel} on the screen, is an infinitestimal area element.   We would
like to specify the orientation of this element in space-time, and
the direction of the null vector penetrating it.   In ancient times,
Cartan and Penrose pointed out that this information is precisely
encoded in a {\it pure spinor}, namely a Dirac spinor which
satisfies

$$\bar{\psi} \gamma^{\mu} \psi \gamma_{\mu} \psi = 0.$$  The null
vector is $\bar{\psi} \gamma^{\mu} \psi $.   The non-vanishing
components of the other Dirac bilinears all lie in a hyperplane
transverse to this null vector, which defines the direction of the
holographic pixel.   The non-vanishing components of the spinor
transform as a spinor of the $SO(d-2)$ tangent space group of the
screen.   Thus, we can think of the orientational information in a
holographic screen as encoded in an element of the spinor bundle
over the screen.   We can quantize the pixel orientation by
postulating commutation relations
   $$[S_a (m), S_b (n) ]_+ = \delta_{mn} \delta_{ab} ,$$ for the
non-vanishing real components of the spinor.   We choose
anti-commutators in order to have a finite dimensional Hilbert space
for each pixel\footnote{One might have expected the degrees of
freedom of independent pixels to commute.   We have made everything
anti-commute by exploiting the $(-1)^F$ gauge invariance of the
formalism\cite{tbprev}.}. This is the most general rule covariant
under the $SO(d-2)$ tangent space group of the holographic screen,
if we restrict each pixel to carry only information about the
orientation of the screen in the non-compact dimensions of
space-time.   Compact factors in the space-time will lead to more
general pixel algebras, and the theory of these is only beginning to
be worked out\cite{tbjf}.

It is remarkable that the algebra of a pixel on the holographic
screen is the same as that of a massless superparticle with fixed
momentum.   This suggests that for very large causal diamonds, our
formalism should describe scattering states of massless
superparticles, and certain examples have been worked
out\cite{11dbfm}, which show that this is kinematically correct. For
consistency, the dynamics must also be supersymmetric.

Returning to our cosmological model, we identify the Hilbert space
${\cal K}$ with the irreducible representation of the single pixel
algebra.  We define a class of random Hamiltonians as follows: First
choose a random fermion bilinear
$$H_2 \equiv \sum S_a (n) A_{mn} S_a (m) .$$   For a large number of
pixels, the spectrum of $H_2$ approaches that of the massless Dirac
equation in $1 + 1$ dimensions, with a cutoff.   We can now add
higher order terms, as long as they are irrelevant perturbations in
the infrared of this $1 + 1$ dimensional field theory.  The spectral
properties of all of these Hamiltonians are the same, except for a
small number of states near the cutoff.

The operators $U(t,k)$ have the form
$$e^{i [H_1 (t,k) + H_2 (t,k)]} , $$ where for each $(t,k)$,
$H_{1,2}$ are chosen independently from the distribution of random
Hamiltonians described above.   They are subject to the restriction
that $H_1$ is built only from those $S_a (n)$ variables that act in
${\cal H} (k) $, while $H_2$ depends only on those which act on its
tensor complement in ${\cal H} (t)$.  We have left the ${\bf x}$
labels off these operators, because our overlap rules specify that
the same random choices are made at each ${\bf x}$.   As far as we
can tell, there is no other way to solve all of our dynamical
consistency conditions.

We define the overlap rule that ${\cal O}(t,t, {\bf x, y}) = {\cal
H} (t - P)$.   The dynamics then satisfies all the overlap
conditions if the sequence of Hamiltonians seen by each observer is
the same.  Given random choices for individual observers, there
seems to be no other way to satisfy consistency.   {\it Thus, our
emergent spatial geometry is forced to be homogeneous by the overlap
consistency conditions.}   This overlap rule also defines an
isotropic spatial metric.  Indeed, the overlap rules in general
define a causality distance on the spatial lattice of observers. The
causal distance between two points at time $t$ is the minimal number
of lattice steps that have to be taken before the overlap Hilbert
space becomes trivial.   Our overlap rule defines a causality
distance which is blind to direction on the lattice.   Thus, the set
of points that are a given causality distance, $R$ from the origin
is the locus of endpoints of self avoiding lattice walks of $R$
steps.   For large $R$ this describes a sphere in Euclidean space,
for any lattice with the topology of Euclidean space.   Once we have
reached the causal boundary at fixed $t$, we continue to define the
distance in terms of the minimal number of lattice steps.   This
definition is flat as well as being homogeneous and isotropic.

I will not go into the details here\cite{holocosm}, but one can also
use the scaling laws of the fermion system at large $N$, to show
that the horizon expands as one would expect for a flat FRW space
with equation of state $p = \rho$.   Furthermore, the conformal
Killing vector of this emergent geometry defines an exact symmetry
of the quantum system in the large $N$ limit.   It is the asymptotic
scaling symmetry of free $1 + 1$ dimensional fermions perturbed by
irrelevant operators.

The energy density in a fixed horizon volume is defined in terms of
the fermion system and the total entropy is of order $N$.   These
relations tell us that the entropy and energy of a horizon volume
satisfy the black hole entropy formula, justifying our assertion
that we have constructed a mathematical model of the dense black
hole fluid.

\subsection{A realistic cosmology}

The mathematical model sketched above is a completely consistent
quantum cosmology, but not one you would want to live in.   All
degrees of freedom inside a horizon volume are, at all times, in
equilibrium.   There can be no particles, no observers separate from
the cosmological background - nothing of interest.   The coarse
grained description of the model as FRW geometry can be derived as
indicated above, but is somewhat misleading.  This is not a geometry
in which anything can propagate.

In this subsection I will describe the heuristic model of more
realistic cosmology, which Fischler and I have developed.   I
emphasize that there is at present no real mathematical model of
this more general system.   The basic idea is to imagine a {\it
normal region} as a defect in the $p = \rho$ background, where
observers on some compact region of the lattice see a more normal
FRW geometry.  We take it to be radiation dominated, with the
thought that the region originally had a black hole in it (to
maximize the entropy of the initial conditions) but that black hole
was too small to merge with the surroundings as the particle horizon
increased.   Instead, a region of empty space is created, and the
black hole evaporates into it, producing a radiation gas.   This
picture is irrelevant to what follows, and should be viewed only as
motivation.

Consider first a spherical region of FRW geometry with equation of
state $p = w \rho$ with $w < 1$, embedded in a $p = \rho$
background. Applying the Israel junction condition to the interface,
we find that the coordinate size of the $w < 1$ region must
shrink.Only in the case $ w = - 1$ can we obtain a stable
configuration of non-decreasing physical size.   Einstein's
equations have solutions corresponding to black holes embedded in
any FRW background.   The horizon of the hole is a marginally
trapped surface of arbitrary size.   If we excise the interior of
the hole and replace it with the static coordinate patch of dS space
with dS and Schwarzschild radii equal, then we satisfy the
longitudinal part of the Israel condition.   The surface stress
tensor implied by the transverse condition, obeys the weak energy
condition.   We conclude that a spherical ball of radiation or
matter dominated universe can only survive if it asymptotes to dS
space and initially occupies a huge coordinate volume, much larger
than our current horizon size.   This seems like a very improbable
initial condition.

In fact, there are surely initial configurations which are more
probable than this, which can also evolve to an asymptotically dS
space with a large radius.   Let us make a \lq\lq tinkertoy ",
connecting together a bunch of balls of normal region, each of
fairly small radius ($< {\it e.g.} 100$) in Planck units. The balls
are glued together into a tinkertoy (Fig. 1) entirely contained in a
ball on the lattice whose coordinate radius is that of the
asymptotic dS horizon.   In describing these balls, we are using the
coarse grained FRW geometry which emerges from the dense black hole
fluid.    The balls describe regions in that background geometry
(which ultimately means a subclass of observer Hilbert spaces
associated with a compact region on the lattice) in which the
dynamics is not what we described in the previous section, but
resembles that of a radiation dominated FRW universe.

Initially, the tinkertoy takes up a small fraction of the physical
volume of the compact region of the lattice corresponding to the
asymptotic dS horizon.   We can define an equal area time slicing on
the inhomogeneous geometry consisting of tinkertoy plus $p=\rho$
background.   At each time, we insist that the area of the causal
diamond that goes all the way back to the Big Bang, is the same at
every lattice point.   In terms of the Hilbert spaces ${\cal H} (t,
{\bf x})$, we are just enforcing the already enunciated rule that
the dimension of all Hilbert spaces at time $t$ is the same.

It is an easy exercise to show that {\it on equal area time slices,}
the volume of space in the $p = {1/3} \rho$ regions grows more
rapidly than that in the dense black hole fluid.   Thus, after some
time, a better picture of the geometry is given by Fig 2. In this
figure we see a large volume normal region, interspersed with small
volumes in which all the degrees of freedom are in equilibrium. It
does not take much imagination to guess that a proper description of
this regime in cosmic history is a dilute black hole gas. We say
that the cosmology has made a transition between the dense black
hole fluid and the dilute black hole gas.  The horizon size at the
time of this transition defines an average black hole mass $M$, and
the energy density is of order ${1\over M^2}$ in Planck units.

If we want to find the most probable initial condition which leads
to this transition, we should look for the tinkertoy which takes up
the smallest fraction of the initial value.  This is because
initials conditions in a normal region are much more constrained
than those in the dense black hole fluid, which saturates the
covariant entropy bound at every moment.  Thus, we expect $M$ to be
a very large number, but we cannot yet calculate it from first
principles.   Note that before this transition it does not make
sense to describe the evolution of the universe in terms of low
energy effective field theory.   The real criterion for the validity
of effective field theory in a space-time region is that the state
of the system is very far from saturating the entropy bound.  The
dense black hole fluid gives us examples of low energy density
configurations which cannot be described by effective field theory.
There is a coarse grained space-time description, but the
excitations of this space-time are not particles traveling through
it, but the low entropy tinkertoys we have just
discussed\footnote{There is something in this discussion reminiscent
of duality transformations in classical statistical mechanics.   In
the Ising model, for example, the maximal entropy high temperature
state is described as a frozen state of dual variables. In some
sense, our tinkertoys are the dual variables with which to define
excitations of the $p = \rho$ system.}.

The tinkertoy configuration evolves to an inhomogeneous state of the
dilute black hole gas, whose inhomogeneities are in principle
determined (statistically) by the most probable choice of tinkertoy
that actually leads to the transition.   We cannot of course figure
out what this is, but one thing is clear.   {\it The fluctuations in
black hole masses around the average value $M$ must be quite small,
and so must the fluctuations in their velocities around a uniform
cosmic velocity.}  Indeed, if this were not so, then the black holes
would quickly collide and merge.   The system would recollapse to
the dense black hole fluid.   While this is qualitatively obvious,
we have not yet been able to turn it into a quantitative bound on
the size of the fluctuations.

The statistics of these fluctuations also depends on what the most
probable configuration for the tinkertoy is and we do not yet have a
grasp of that.   However, the exact scale invariance of the $p=\rho$
system enables us to show that the two point function of these
fluctuations satisfies Harrison-Zeldovitch scaling in a range of
scales whose physical size runs from the Planck scale to the size
$M$ of the horizon at the transition to the dilute black hole gas
phase.  The fluctuations outside this range are cut off. Recent
observations seem to suggest that such exact scale invariance is
ruled out for fluctuations in the CMB in the largest few orders of
magnitude of scales on the sky.

Another problem with the model so far is that the correlation length
of these fluctuations cannot be as big as that of the correlations
in the CMB unless we have a period of inflation.   This is the {\it
real} horizon problem. Only inflation can explain correlations in
the fluctuations on our current particle horizon scale. We have seen
that approximate homogeneity and isotropy follows from the
properties of the dense black hole fluid, combined with the
requirement that the normal regions do not recollapse into the $p
=\rho$ phase.   Note however that we only need 10-20 e-folds of
inflation to explain the observed correlations in the CMB.

After the phase transition to the dilute black hole gas, we can
begin to use effective field theory to describe the dynamics of the
universe.   The detailed microstates of the black holes are not well
described by this formalism, but their coarse grained thermodynamics
is.   This effective field theory might be fairly unique, or it
might be part of a landscape.  However, our cosmology terminates in
a stable dS space with a finite number of states\footnote{This is a
probabilistic argument.   The String Landscape assumes that the
history of the universe has an infinite number of states that are
well described by low energy effective field theory.   Even if it
exists, it would seem to be a less probable way for the universe to
escape from the dense black hole fluid than the one we are
proposing.}.   Thus, the potential on the Landscape should be above
the Great Divide\cite{abj} with regard to transitions out of the dS
minimum with lowest c.c. .   At any rate, we will assume that the
potential on the space of low energy scalar fields allows for
inflation.

Recall that for potentials which vary in field space over field
intervals of order the reduced Planck scale $m_P$, we only need a
$1\over N_e$ coincidence to have $N_e$ e-folds of inflation.
Furthermore potentials of this form are technically natural if all
the couplings of the relevant scalar fields are suppressed by $m_P$.
Finally, the initial conditions for the field {\it must} be fairly
homogeneous, in order to avoid recollapse into the dense black hole
fluid.   It is reasonably probable that the initial field
configuration will be sufficiently homogeneous for inflation to
begin, as long as the inflationary energy density is well below
$1/M^2$, so that we can be sure that we are in a regime described by
low energy effective field theory.   This probably constrains the
potential to be such that no measurable tensor fluctuations will be
generated during inflation. We\cite{tbwfunpub} have searched for
other mechanisms, which could generate a spectrum of primordial
gravitational waves in holographic cosmology, but have not found any
so far.   Thus, one could tentatively conclude that holographic
cosmology predicts a vanishing tensor to scalar ratio.  The most
important conclusion is that holographic cosmology provides an
explanation of the homogeneous initial conditions necessary for
inflation to work.

The amount of inflation necessary to explain the data in this model
is definitely smaller than in garden variety inflation models. Gross
flatness, homogeneity and isotropy are explained by a
non-inflationary mechanism. Inflation is only necessary in order to
explain horizon spanning correlations, though it can also be the
source of the CMB fluctuations.  The necessary amount of inflation
depends on the value of $M$, the scale of inflation, and the
question of whether we want to explain observations in the CMB with
inflationary fluctuations, or fluctuations generated during the
$p=\rho$ era. Preliminary observations, which suggest that the
spectrum is not {\it exactly} scale invariant, favor the
inflationary fluctuations. In any case, we do not need more than
$30$ e-folds of inflation.

Since my purpose here is not to delve into the details of
holographic cosmological models, I will stop, and reiterate the most
important point of this section.   Holographic cosmology provides an
explanation for the apparent low entropy of cosmological initial
conditions.   The initial conditions are the highest entropy
possible, consistent with avoiding the collapse into a dense black
hole fluid.   We have provided qualitative arguments for this, but
not a quantitative derivation of the degree of homogeneity we
observe in the universe.   As a bonus we predict that the final
state of the universe is asymptotically dS with an {\it a priori}
distribution for the c.c., which favors large values.

The DKS model explains low entropy initial conditions as an unlikely
fluctuation, and appears to founder on the bizarre phenomenon of
Boltzmann's brain.   The model has a history which parallels the
cosmology we have discovered by observation, but seems to predict
that a typical observer will instead arise as a local thermal
fluctuation.   Holographic cosmology does not suffer from this
problem.   Its typical cosmological history goes through a
conventional evolutionary production line for observers, with
conditional\footnote{The only condition imposed is that the universe
does not evolve forever as a dense black hole fluid.} probability
one. The fact that it also predicts rare Boltzmann Brain events in
the distant future is no more problematic than the fact that we have
never seen all the air in the room spontaneously collect in a
corner.

\section{Conclusions }

We have reanalyzed the model of DKS, which provides a reason for
cosmological evolution to begin in a low entropy state. The basic
mechanism for this is a Poincare recurrence of a low entropy
fluctuation in a finite entropy system.   This is an old idea,
possibly going back to Boltzmann, but models of this type generally
run into a paradox. If we imagine that cosmological evolution is the
same unitary groupoid as the one which generates the recurrences, we
can argue that the typical recurrence of cosmology, even when
subjected to anthropic constraints, does not resemble the one we
see.

We argued that the way out of this paradox comes in the recognition
that the time dependent Hamiltonian, $P_0 (t)$ which describes
cosmology, is {\it not the same} as the static Hamiltonian, $H$,
which describes the recurrences.   $P_0 (t)$ is instead an emergent
operator, describing the evolution of a small subset of the degrees
of freedom of the system, over time periods much shorter than the
recurrence time of $H$.   Much of interesting cosmological history
in fact occurs over times short compared to the asymptotic Hubble
time, $R$.    We argued that recurrences of events where $P_0 (t)$
is a relevant description of a subset of the system all resemble
each other apart from microscopic details, and the description of
events in the very far future\footnote{In this model, most of the
visible universe will go out of our horizon in a few times the
current age of the universe. Even our gravitationally bound
environment will eventually be destroyed by gravitational collapse
or dS thermal fluctuations .   It is likely that the history of
these very late times, and particularly the effect of dS
fluctuations, can recur in very different ways. Even if we should
live so long we might not be able check whether the very late cosmic
history we see is a typical member of the ensemble of all histories.
Such a check would perforce involve knowledge of what has happened
to parts of the universe outside our horizon. Our asymptotically dS
universe has of order $(RM_P)^{1/2} \sim 10^{30} $ independent
horizon volumes at late times, so we can expect to see coincidences
of relative probability $10^{ - 30}$, if we only examine localized
excitations in our own horizon volume. }. The model simply does not
contain cosmological histories with {\it e.g.} different cosmic
microwave background temperatures.

The last statement is confusing to the effective field theorist,
because at late times the CDL cosmology appears to allow such
states.    However, if we extrapolate this late time data back, we
find a space-like singularity intervening between us and the
coordinate singularity in the CDL solution (the bubble nucleation
event).    Thus, there is no way of connecting this data to any
controlled fluctuation of the asymptotic dS vacuum state\footnote{Of
course, if one were perverse one could insist on including all of
these late time states as part of the evolution of the system, and
proceed to rederive the DKS paradox.    My point is that the model
provides us with no evidence that these states have anything to do
with fluctuations of a finite system.   Indeed, the form of the CDL
amplitudes, which obey the principle of detailed balance, suggests
that the entropy of states which have anything to do with the actual
tunneling event is much smaller than what one would estimate with
cutoff field theory at late times. Cosmological evolution of these
low entropy initial states, produces states which are
indistinguishable (after a little coarse graining) from a much
higher entropy ensemble.   Most of the members of that late time
ensemble also extrapolate back to a singularity, but we are not
concerned with this because their prediction for the future are hard
to distinguish from those of the initial low entropy ensemble.
However, states which are distinguishable, like those with higher
than expected CMB temperature, cannot be connected to any well
established fluctuation of the system. }. The effective field theory
in CDL cosmology is only valid for a small subset of the possible
states in that field theory at late times. The other states of that
effective theory are likely to have nothing to do with the correct
description of any event in the full system. There are many
fluctuations of the late time dS geometry which do not resemble
possible events in the CDL cosmology, but there is no reason to
imagine that they resemble anything close to it at all. For example,
the late time dS space has black holes of size much larger than the
size of the critical bubble.   These are certainly produced as
thermal fluctuations in the future, but have no description in terms
of the degrees of freedom which follow the evolution of $P_0 (t)$. A
microscopic description of the events leading to the nucleation of
such black holes involves the dynamics of the Hamiltonian $H$ in a
crucial way, although we can estimate the probability of these
events by doing the statistical mechanics of $P_0$ at the dS
temperature.

A residual problem for DKS cosmologies is the thermal production of
small intelligent systems.   We argued that this could be a problem
for the string landscape as well.  It is likely that, if it exists,
the landscape contains approximately SUSic dS spaces. If any of
these have low energy physics resembling our own, then the string
landscape will have Boltzmann brains.   It is not possible at this
time to decide whether these are the most probable form of
intelligent life in the string landscape.   This is directly
connected to the fact that no one has addressed the problem of
initial conditions and the explanation for their low entropy, in the
string landscape context. The conventional wisdom is that we get
into the basin of attraction of our minimum by CDL tunneling, which
could provide an explanation for homogeneity, isotropy and slow roll
inflation.   However, unlike the DKS model, the landscape provides
no way of estimating the probability of getting into the low entropy
false dS vacuum which begins the current era of history of the
universe.

Finally we reviewed the way in which Holographic Cosmology attempts
to explain the low entropy of cosmological initial conditions, and
the way in which it avoids the problem of Boltzmann brains. BBs
exist, but are less probable ways to produce intelligence than that
afforded by conventional cosmology and evolution.  {\it Indeed,
there is a way to state our conclusions about holographic cosmology
which is quite general.   If cosmology is described by quantum
mechanics with a time dependent Hamiltonian, then in order to avoid
the problem of Boltzmann's Brain we need the space of initial states
to divide into two categories.   In the first category, life (or
perhaps just life of our type)is not possible.   In the second, the
universe undergoes a period of cosmological expansion with time
dependent Hamiltonian and produces life via more or less standard
cosmology and evolution. In the second category the far future
evolution of the universe may produce a large number of Boltzmann
brains, if the future is described by a stable dS space.   This is,
in my opinion, no more problematic than the fact that the ordinary
laws of thermodynamics say that if a room exists long enough, all
the air in it will collect in a corner.   Every history in our model
that supports Boltzmann brains also produces a prior period of
evolved observers.   This period never repeats as a thermal
fluctuation, because the Hamiltonian describing the ultimate
equilibrium state of the universe is not the time dependent
Hamiltonian describing cosmology.  This is similar to our discussion
of the DKS model, but in the present class of model evolved life
occurs with probability one.

Boltzmann Brains, and their support systems, can be observed by
evolved observers, only if their size is much smaller than the
cosmological horizon.   Otherwise, the waiting time for producing
the BB support system is longer than the maximal survival time of a
local observer\cite{nightmare}.   Thus, in this class of models, BBs
are properly viewed as the kind of crazy fluctuation statistical
mechanics teaches us to expect and discount.   Any observer can tell
whether it is a BB by making a relatively small number of
observations.   If our own observations are described by such a
model, then it is certain that we are not BBs.}

If one wants to believe the mathematical predictions of this class
of models, over time scales much longer than the maximal survival
time of an evolved local observer, then fluctuations might occur,
which reproduce the current state of all localizable degrees of
freedom in our universe.    One could then choose to evolve that
state with the time dependent Hamiltonian $P_0 (t)$ starting at the
present time and extending for times short compared to ${1\over
R_{dS}}$.   During this time period the actual evolution under $H$,
which is prescribed by the definition of the model, does almost
nothing to the state and so it makes sense to study an emergent
Hamiltonian which describes variations over shorter time
scales\footnote{Though it seems completely arbitrary to use exactly
the same Hamiltonian $P_0(t)$ that the model prescribed for an
earlier period of the universe.}.   This kind of BB would be hard to
distinguish from ourselves, but it is enormously less probable than
the smaller BBs.    So, given a model of this type, we would
conclude from our observations that we were either the original,
evolved, observers, which are guaranteed to exist in any
anthropically allowed history of the model, or an enormously
improbable thermal fluctuation.   I leave it to the reader to decide
which explanation we would be likely to choose.

The idea that the second law of thermodynamics could be explained by
assuming that our cosmological history was a fluctuation in a finite
system probably originates with Boltzmann.   His vast intelligence
still haunts our considerations of this proposal.  The other class
of models described above, exemplified by holographic cosmology,
provides a more satisfactory understanding of the apparent low
entropy of cosmological initial conditions.

\section{Acknowledgments}

I would like to thank Willy Fischler, Lorenzo Mannelli and Tomeu
Fiol for contributing to the suite of ideas that formed the basis of
this paper. I'd also like to thank A.Aguirre, M.Johnson, A. Linde,
B.Freivogel, M.Kleban and L.Susskind for comments on the manuscript.
I particularly want to thank R. Bousso, for detailed criticism of a
previous version of this manuscript.

This research was supported in part by DOE grant number
DE-FG03-92ER40689.

\end{document}